\documentclass[
  amsmath,
  amssymb,
  preprint
]{revtex4-2}

\usepackage{graphicx}
\usepackage{dcolumn}
\usepackage{bm}
\usepackage{svg}

\usepackage[mathlines]{lineno}

\makeatletter
\newcommand{\nonfloattablecaption}{\def\@captype{table}\caption}
\newcommand{\nonfloatcaption}{\def\@captype{figure}\caption}
\makeatother

\begin{document}

\title[]{Cascade-driven fragmentation-aggregation transitions in decaying turbulence}

\author{Vivek Kumar}
\author{Prasoon Suchandra}
\author{Shivam Prajapati}
\author{Suhas Jain}
\author{Cyrus Aidun}

\affiliation{%
  \parbox[t]{\textwidth}{%
    \centering
    \footnotesize
    \vspace*{6mm}%
    George W. Woodruff School of Mechanical Engineering, Georgia Institute of Technology, USA\\[2mm]
    Daniel Guggenheim School of Aerospace Engineering, Georgia Institute of Technology, USA\\[2mm]
    Center for Multiphase Flow Research, Georgia Institute of Technology, USA%
  }%
}

\begin{abstract}
In decaying turbulence, the turbulent cascade is not merely a passive background for breakup but an evolving stability boundary that dynamically selects the fragmentation–aggregation pathway. As turbulence decays, the Hinze scale sweeps through the bubble-size distribution, progressively eliminating the breakable population and driving a transition from a mixed breakup–coalescence regime to a self-sustaining pure-coalescence state. Our model predicts a critical  turbulence decay exponent \(m_c=5/3\), a transient mixed breakup--coalescence regime, and a late stage pure-coalescence state with a universal coalescence hazard \(r(t)\sim t^{-1}\). Interestingly, the transient breakup acts as a catalyst that amplifies mixed-regime bubble growth relative to pure coalescence. Results from our direct numerical simulations of decaying homogeneous isotropic turbulence as well as high-Reynolds-number duct-flow experiments collapse onto the model predictions when expressed in terms of the measured decay rate. These results replace the conventional instantaneous Hinze criterion with a non-equilibrium, cascade-controlled framework for turbulent fragmentation and aggregation.
\end{abstract}

\date{\today}

\maketitle

\begin{figure}[htbp]
\includegraphics[width=0.8\textwidth]{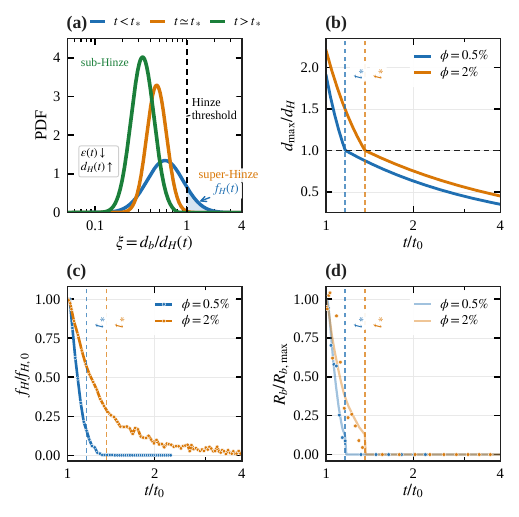}
\caption{Dynamic suppression of breakup by Hinze-scale drift.
(a) Schematic bubble-size PDF relative to the instantaneous Hinze scale, $\xi=d_b/d_H(t)$, showing depletion of the super-Hinze tail.
(b) Robust upper-tail diameter normalized by the Hinze scale,
\(d_{\max}(\simeq d_{99.8})/d_H\), used to identify the transition into the
sub-Hinze regime.
(c) Normalized DNS super-Hinze fraction, $f_H/f_{H,0}$, for $\phi=0.5\%$ and $2\%$.
(d)  Normalized resolved-scale breakup contribution, \(R_b/R_{b,\max}\), computed
from resolved breakup events, showing that breakup becomes negligible after the transition. Dashed vertical lines mark $t_\ast=7.1$ and $8.3$, corresponding to the transition times for $\phi=0.5\%$ and $2\%$, respectively.
Time is normalized by $t_0=6.1$.}
\label{fig:f_H}
\end{figure}

\textit{Introduction}-- Dispersed drops and bubbles in turbulence mediate transport, mixing, and interfacial exchange in natural and engineered flows, from clouds and air--sea gas transfer to sprays, emulsions, and bubbly reactors~\cite{Mathai2018Dispersion,Lohse2018BubblePuzzles}. Their size distribution is set by a competition between fragmentation, which increases the number of dispersed elements, and aggregation or coalescence, which reduces their number and increases their characteristic size. The standard picture of turbulent fragmentation is local and instantaneous: inertial stresses associated with turbulent velocity differences at the dispersed-phase scale are balanced against stabilizing interfacial or cohesive stresses, defining a critical stability scale, such as the Kolmogorov--Hinze scale, above which breakup is expected~\cite{Kolmogorov1949,Hinze1955,Xu2024Intermittency,MasukSalibindlaNi2021Hinze,QiTanCorbittUrbanikSalibindlaNi2022SmallEddies}. This framework has shaped much of our understanding of breakup in turbulent flows, but it treats the turbulent cascade as statistically stationary. In freely decaying turbulence, however, the cascade is itself time dependent: the dissipation rate decreases, the Hinze scale grows, and the dispersed population simultaneously evolves through breakup and coalescence. The fragmentation--aggregation balance is therefore not determined solely by an instantaneous Weber-number threshold, but by the relative rates at which the turbulence and the dispersed phase evolve~\cite{RivierePerrard2025BreakupProbability}. This raises a basic question: can turbulent decay itself suppress fragmentation and drive a transition to aggregation-dominated dynamics?\\
In decaying homogeneous isotropic turbulence (HIT), the cascade evolution is constrained by the large-scale structure of the flow. For low-wavenumber spectra of the form $E(\kappa)\sim \kappa^{p-1}$ as $\kappa\to0$, the classical Saffman and Loitsyanskii--Batchelor invariants give a power-law decay of the dissipation rate, $\varepsilon(t)\sim t^{-m}$, with $m=(3p+2)/(p+2)$; thus $m=2.2$ for Saffman turbulence $(p=3)$ and $m\simeq2.43$ for Loitsyanskii--Batchelor turbulence~\cite{saffman1967large,batchelor1956large,loitsyanskii1939some,Sinhuber2015Decay,JohnDonzisSreenivasan2022DecayLaws} $(p=5)$. High-Reynolds-number measurements show that observed decay laws and inertial-range statistics can contain finite-Reynolds-number and large-scale-structure corrections~\cite{Sinhuber2015Decay,Sinhuber2017Dissipative,Kuchler2023Universal,Iyer2021Oscillations,JohnDonzisSreenivasan2022DecayLaws,MeldiSagautLucor2011StochasticDecay}; hence, we treat $m$ as a measured property of the decaying cascade rather than as a universal constant. Under the dilute conditions considered here, $\phi=\mathcal{O}(10^{-2})$, the
dispersed-phase feedback remains weak enough that the bulk turbulent decay
stays close to the carrier-flow cascade, as quantified in the Supplemental
Material. Breakup and coalescence, however, remain controlled by bubble-scale
interfacial dynamics~\cite{Mathai2018Dispersion,Tan2025PairDispersion,Lvov2005Drag,Verschoof2016Drag,Ma2025Kolmogorov}. The key consequence is that the Hinze scale experienced by the dispersed phase evolves with the cascade, allowing the leading-order population exponents to be predicted once the decay exponent $m$ is specified.\\
In this letter, we ask how a time-dependent turbulent cascade selects the pathway from fragmentation to aggregation. We develop a theoretical population-balance-type framework in which the breakable fraction is coupled to the evolving Hinze scale, while bubble-bubble encounters are described by inertial-range collision dynamics. The framework is tested using direct numerical simulations (DNS) of dilute bubbly decaying HIT and high-Reynolds-number duct-flow experiments. The numerical setup, experimental facility, and measurement procedures are detailed in Ref.~\cite{vivek_jfm}. Together, the DNS and experiments provide complementary realizations of idealized and spatially developing decaying turbulence. Rather than treating the Hinze criterion as a fixed instantaneous cutoff, the analysis follows the relative evolution of the bubble population and the Hinze scale, thereby identifying when breakup remains active, when it is suppressed, and how coalescence-dominated dynamics emerge. Fig.~\ref{fig:f_H} summarizes this transition.
Fig.~\ref{fig:f_H}(a) illustrates the Hinze-scale drift through the bubble-size PDF.
Fig.~1(b) shows the upper-tail crossing,
\(d_{\max}/d_H\simeq d_{99.8}/d_H\), where \(d_{99.8}\) denotes a statistically
converged upper-tail diameter evaluated from the largest \(0.5\%\) of bubbles,
following Ref.~\cite{kumar2026bubble}. This crossing identifies when the measurable
breakable tail falls below the Hinze scale. Figs.~\ref{fig:f_H}(c)
and~\ref{fig:f_H}(d) show the associated decay of the super-Hinze fraction
\(f_H/f_{H,0}\) and the resolved breakup contribution
\(R_b/R_{b,\max}\), respectively.

\textit{Model, observables, and results}: We describe the dispersed phase by the number density \(n(t)\), representative
diameter \(d(t)\), and number-weighted bubble-size PDF \(p(d_b,t)\). Starting from the population-balance equation for binary coalescence and turbulent breakup~\cite{Coulaloglou1977,ramkrishna2000population,Lehr2002,ChanJohnsonMoin2021CascadePart1,Leyvraz2003AggregationReview}, and integrating over bubble size, the number balance reduces to
\begin{equation}
\dot{n} = -\frac{1}{2}\Gamma n^2
+ \mathcal{B} f_H(t)n ,
\label{eq:pbe_reduced_main}
\end{equation}
where an overdot denotes differentiation with respect to time, $\Gamma$ is the effective coalescence kernel, $\mathcal{B}$ is the breakup frequency, and
\begin{equation}
f_H(t) = \int_{d_H(t)}^\infty p(d_b,t)\,\mathrm{d}d_b
\label{eq:fH_main}
\end{equation}
is the fraction of bubbles larger than the instantaneous Hinze scale $d_H(t)$. The first term in Eq.~\eqref{eq:pbe_reduced_main} removes one bubble per coalescence event, while the second term represents the net production of bubbles by breakup. The resolved-scale breakup contribution shown in Fig.~\ref{fig:f_H}(d) is computed from tracked breakup events; its reduced-model counterpart is \(R_b=\mathcal{B}f_Hn\). The detailed steps of reduction from the full population balance, including binary breakup, volume conservation, and the effective-monodisperse approximation, are provided in the Supplemental Material.

The Hinze scale follows from the turbulent Weber-number balance at the bubble scale~\cite{Kolmogorov1949,Hinze1955,Xu2024Intermittency}. For decaying turbulence with $\varepsilon(t)\sim t^{-m}$,
\begin{equation}
    d_H(t)\sim \varepsilon(t)^{-2/5}\sim t^{2m/5}.
\end{equation}
Thus, unlike in stationary fragmentation theories, the breakup threshold is not fixed; it changes through the evolving bubble-size distribution. The bubble-bubble collision rates are closed using inertial-range estimates. For bubbles of comparable size, the relative approach velocity for bubble diameter $d$ is $u_d\sim(\varepsilon d)^{1/3}$, giving the encounter kernel~\cite{batchelor1956large,moninm,prince1990bubble,luo1996breakup,JiangKrug2025CollisionRate} $h\sim u_d d^2\sim \varepsilon^{1/3}d^{7/3}$. The effective coalescence kernel is $\Gamma=h\lambda$, where $\lambda$ is the coalescence efficiency. The widely accepted film-drainage model shows that $\lambda$ depends on the
ratio of the film-drainage time to the contact time between bubbles
~\cite{chesters1991modelling,liao2009review,Liu2019Coalescence,chan2011film}. This ratio depends on both the turbulent dissipation rate and the bubble size. In the present highly turbulent decaying flow, the decrease in dissipation is partly compensated by the monotonic growth of the bubble size, so the ratio
varies very weakly with time and $\lambda$ remains high and nearly constant~\cite{kumar2026bubble}. The breakup frequency is estimated from the inverse eddy-turnover time at the bubble scale~\cite{MartinezBazan1999,garrett2000connection,Riviere2021,ChanJohnsonMoin2021CascadePart1,ZhongNi2024BreakupFrequency}, $\mathcal{B}\sim\tau_d^{-1}\sim\varepsilon^{1/3}d^{-2/3}$. With approximately conserved gas volume, $\phi\sim nd^3$, so that $n\sim d^{-3}$. Substitution into Eq.~\eqref{eq:pbe_reduced_main} gives, up to order-unity prefactors,
\begin{equation}
\dot{d}
\sim
\varepsilon^{1/3}d^{1/3}
\left[\frac{1}{2}-f_H(t)\right].
\label{eq:mixedregime}
\end{equation}
Bubble growth is therefore controlled by both the turbulent encounter rate and the depletion of the breakable population.

During the mixed coalescence--breakup regime, the bubble-size distribution still spans over the Hinze scale. A local expansion of $f_H(t)$ near $d_H(t)\simeq d(t)$ gives
\begin{equation*}
\frac{1}{2}-f_H(t)
\sim
\frac{d_H(t)}{d(t)}-1 ,
\end{equation*}
as derived in the Supplemental Material. Assuming $d(t)\sim t^{\beta_t}$ and using Eq.~\eqref{eq:mixedregime}, the leading-order transition-regime exponent becomes
\begin{equation}
\beta_t=\frac{15+m}{25}.
\label{eq:beta_mixed_main}
\end{equation}

This exponent is not an asymptotic law for all times; it applies while the Hinze scale remains within the active support of the bubble-size distribution. In this regime, $n(t)\sim t^{-3\beta_t}$, $A(t)\sim nd^2\sim t^{-\beta_t}$, and the per-bubble coalescence hazard $r(t)\equiv\Gamma n$ scales as $r(t)\sim t^{-m/3-(2/3)\beta_t}$.

\begin{center}
\small
\setlength{\tabcolsep}{5pt}
\renewcommand{\arraystretch}{1.15}

\resizebox{\linewidth}{!}{%
\begin{tabular}{lccccc}
\hline
\textbf{Quantity}
& \textbf{Mixed regime}
& \multicolumn{2}{c}{\textbf{Mixed-regime exponent}}
& \shortstack{\textbf{Pure coalescence}\\\textbf{regime}}
& \shortstack{\textbf{Pure coalescence}\\\textbf{regime exponent}} \\
\cline{3-4}
& & $m_s=2.2$ & $m_b=2.43$ & & $m_s=2.2,\;m_b=2.43$ \\
\hline

$d(t)$
& $\sim t^{\beta_t}$
& $0.688$ & $0.697$
& $\sim t^{(3-m)/2}$
& $0.40,\;0.28$ \\

$a_b(t)$
& $\sim t^{2\beta_t}$
& $1.376$ & $1.394$
& $\sim t^{3-m}$
& $0.80,\;0.57$ \\

$A(t)$
& $\sim t^{-\beta_t}$
& $-0.688$ & $-0.697$
& $\sim t^{(m-3)/2}$
& $-0.40,\;-0.28$ \\

$n(t)$
& $\sim t^{-3\beta_t}$
& $-2.064$ & $-2.092$
& $\sim t^{-(9-3m)/2}$
& $-1.20,\;-0.85$ \\

$\Gamma(t)$
& $\sim t^{-m/3+(7/3)\beta_t}$
& $0.872$ & $0.817$
& $\sim t^{(7-3m)/2}$
& $0.20,\;-0.15$ \\

$r(t)$
& $\sim t^{-m/3-(2/3)\beta_t}$
& $-1.192$ & $-1.275$
& $\sim t^{-1}$
& $-1.00,\;-1.00$ \\

$R(t)$
& $\sim t^{-m/3-(11/3)\beta_t}$
& $-3.256$ & $-3.366$
& $\sim t^{(3m-11)/2}$
& $-2.20,\;-1.85$ \\

$\mathcal{B}(t)$
& $\sim t^{-m/3-(2/3)\beta_t}$
& $-1.192$ & $-1.275$
& --
& -- \\

$d_H(t)$
& $\sim t^{2m/5}$
& $0.880$ & $0.972$
& $\sim t^{2m/5}$
& $0.88,\;0.97$ \\

$d(t)/d_H(t)$
& $\sim t^{\beta_t-2m/5}$
& $-0.192$ & $-0.275$
& $\sim t^{(15-9m)/10}$
& $-0.48,\;-0.69$ \\

$t_\ast$
& $\sim \chi_i^{25/(9m-15)}$
& $\chi_i^{5.208}$ & $\chi_i^{3.639}$
& --
& -- \\

\hline
\end{tabular}
}

\nonfloattablecaption{Temporal scaling of key bubble-population and
multiphase quantities in decaying HIT. The mixed regime corresponds
to the transition regime where coalescence and breakup coexist,
with $d(t)\sim t^{\beta_t}$ and $\beta_t=(15+m)/25$. The pure
coalescence regime corresponds to the asymptotic sub-Hinze regime,
with $d(t)\sim t^{(3-m)/2}$. Exponents are evaluated for representative
decay rates $m=2.2$ and $m=2.43$. Here $m_s=2.2$ and $m_b=2.43$
denote representative Saffman and Loitsyanskii--Batchelor decay
exponents, respectively. The quantity $a_b$ is the interfacial area
of a single bubble, $A$ is the interfacial area per unit volume,
$\Gamma$ is the coalescence kernel, $r=\Gamma n$ is the per-bubble
coalescence hazard, $R=\Gamma n^2$ is the volumetric coalescence
rate, and $\mathcal{B}$ is the breakup frequency. Here $t_\ast$ is
the transition time to the pure-coalescence regime,
$\chi_i=d_{99.8}(t_i)/d_H(t_i)$ is the initial upper-tail ratio,
$t_i$ is the initial time at which this ratio is evaluated, and
$t_v$ is the virtual origin of the turbulent decay.}
\label{tab:merged_scalings}

\end{center}

As the turbulence decays, \(d_H(t)\) grows and the super-Hinze fraction \(f_H(t)\), defined in Eq.~(2), decreases, providing a direct measure of the breakable population. For the HIT DNS perfoemed at $\phi=0.5\%$ and $\phi=2\%$, as shown in Fig.~\ref{fig:f_H}, \(f_H\) decreases rapidly during the mixed breakup--coalescence regime and follows an effective finite-window scaling
\(f_H(t)\sim t^{-3}\). This exponent should not be interpreted as a universal asymptotic law, but as an effective depletion rate of the initially populated super-Hinze tail. This rapid depletion is consistent with two DNS observations. First, during the initial cascade-adjustment period, the dissipation decay is steeper than the canonical Saffman or Loitsyanskii--Batchelor values, with a local effective exponent close to \(m_{\rm eff}\simeq3\), as reported in Ref.~\cite{vivek_jfm}.

Since \(d_H\sim\varepsilon^{-2/5}\), this gives \(d_H\sim t^{6/5}\), allowing the Hinze threshold to sweep rapidly through the bubble-size distribution. Second, the initial DNS distribution is non-Gaussian: the sub-Hinze range follows approximately \(p(d_b)\sim d_b^{-3/2}\), while the super-Hinze tail follows the steeper scaling $p(d_b)\sim d_b^{-10/3}~$\cite{ChanJohnsonMoinUrzay2021CascadePart2,RuthAiyerRivierePerrardDeike2022SubHinze,MostertPopinetDeike2022BreakingWaves}. For a normalized super-Hinze tail \(p(d_b)\propto d_b^{-q}\), with its amplitude set by the characteristic diameter, \(f_H\sim(d/d_H)^{q-1}\); therefore the observed \(q=10/3\) tail gives \(f_H\sim(d/d_H)^{7/3}\). For a finite measurable cutoff \(d_u(t)\) of the super-Hinze tail, the same
integration gives an additional factor \(1-(d_H/d_u)^{q-1}\), which vanishes
as \(d_u\to d_H\). Thus, rapid growth of \(d_H\) and depletion
of the finite upper tail make the observed finite-window decay of
\(f_H\) much faster than the infinite-tail estimate. Although \(f_H\) remains finite at late times, no resolved breakup events are detected once the absolute super-Hinze fraction falls below \(f_H\lesssim0.02\) for \(\phi=0.5\%\) and \(f_H\lesssim0.03\) for \(\phi=2\%\). The larger residual \(f_H\) at \(\phi=2\%\) reflects more frequent intermittent super-Hinze excursions generated by coalescence between large bubbles, rather than a sustained breakable population. This shutdown is reflected directly in Fig.~1(d), where the normalized resolved breakup contribution collapses after \(t_\ast\). Because very small daughter bubbles may be under-resolved, the resolved breakup statistics are used as a supporting diagnostic, while the transition is defined primarily by depletion of the super-Hinze tail and the upper-tail condition \(d_{99.8}<d_H\). We therefore define the transition time \(t_\ast\) as the onset of the regime in which breakup becomes negligible and the measurable upper tail lies below the Hinze scale, \(d_{99.8}<d_H\). For the present DNS, this gives \(t_\ast\simeq7.1\) for \(\phi=0.5\%\) and \(t_\ast\simeq8.3\) for \(\phi=2\%\).\\
For \(t>t_\ast\), Eq.~\eqref{eq:pbe_reduced_main} reduces to the pure-coalescence balance \(\dot{n}=-(1/2)\Gamma n^2\). 
Equivalently, \(\dot{d}\sim\varepsilon^{1/3}d^{1/3}\), which gives
\begin{equation}
d(t)\sim t^{(3-m)/2}.
\label{eq:pure_diameter_scaling_main}
\end{equation}
The corresponding population statistics are $n(t)\sim t^{-3(3-m)/2}$ and $A(t)\sim t^{(m-3)/2}$. The coalescence hazard becomes
\begin{equation}
r(t)=\Gamma n\sim t^{-1},
\label{eq:hazard_universal_main}
\end{equation}
independent of $m$.

\begin{figure}[htbp]
    \includegraphics[width=0.8\textwidth]{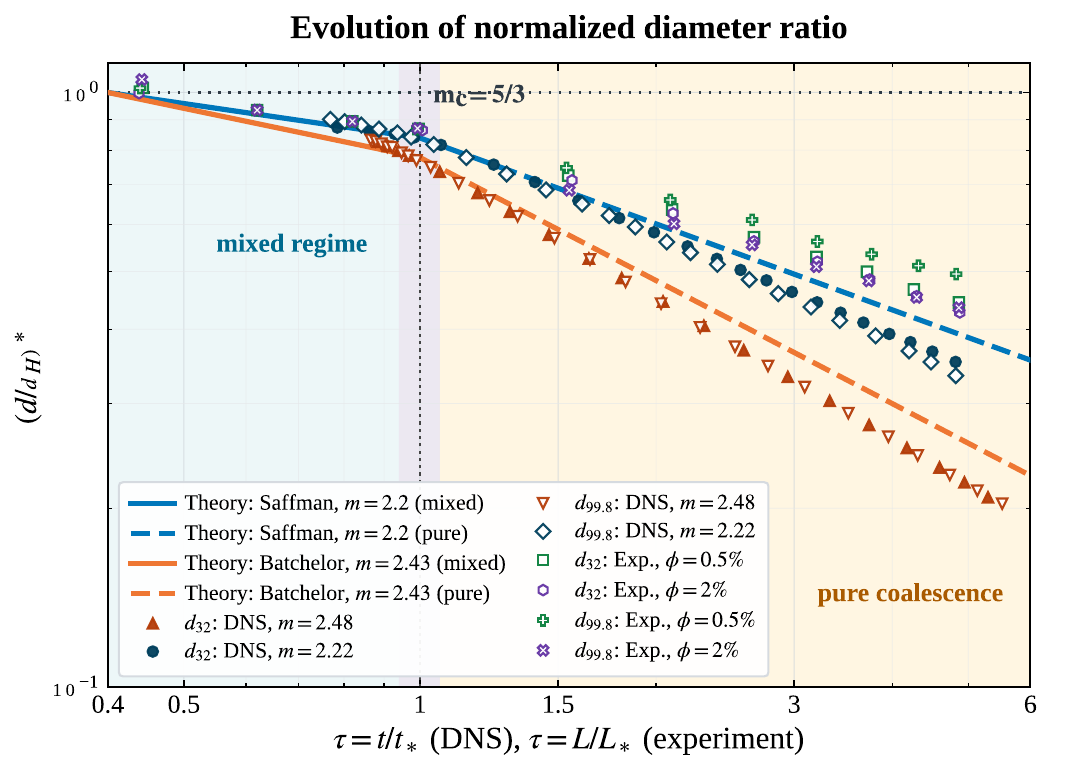}
    \caption{Evolution of the normalized characteristic bubble size as a function of normalized time, showing cascade-driven suppression of fragmentation. The Sauter mean diameter $d_{32}$ and the upper-tail diameter $d_{99.8}$ are normalized by the instantaneous stability scale $d_H$, and time is normalized by the transition time, $\tau=t/t_\ast$, where $t_\ast$ denotes the onset of the pure-coalescence regime. Solid and dashed curves represent theoretical predictions for Saffman ($m=2.2$) and Loitsyanskii--Batchelor ($m\simeq2.43$) turbulence, corresponding to low-wavenumber energy spectra $E(\kappa)\sim \kappa^{p-1}$ with $p=3$ and $p=5$, respectively. Symbols denote DNS and experiments in a decaying duct flow ($m\simeq 2.1$-$2.48$). The critical condition $m_c=5/3$ separates marginal behavior, for which $d/d_H$ remains constant, from the present cases with $m>m_c$, for which $d/d_H$ decreases. For all cases with $m>m_c$, $d_{32}/d_H$ decreases monotonically, demonstrating that the growth of the stability scale outpaces the growth of the dispersed phase, leading to a dynamical suppression of fragmentation. The vertical line at $\tau=1$ marks the transition from a mixed regime, where breakup and coalescence coexist, to a pure coalescence regime.}
    \label{fig:d/d_H}
\end{figure}

The suppression of breakup follows from the relative drift between the bubble population and the Hinze scale:

\begin{equation}
\frac{d(t)}{d_H(t)}
\sim
\begin{cases}
t^{(15-9m)/25}, & t<t_\ast \quad \text{mixed regime},\\
t^{(15-9m)/10}, & t>t_\ast \quad \text{pure coalescence}.
\end{cases}
\label{eq:diameter_hinze_drift_main}
\end{equation}

The absolute transition time also depends on the initial upper-tail ratio and on the virtual origin of the turbulent decay. Writing the decay as $\varepsilon(t)\sim(t-t_v)^{-m}$ and defining the transition by the disappearance of the measurable breakable tail, $d_{99.8}(t_\ast)/d_H(t_\ast)=1$, the mixed-regime drift gives
\begin{equation}
\frac{t_\ast-t_v}{t_i-t_v}
\sim
\left[
\frac{d_{99.8}(t_i)}{d_H(t_i)}
\right]^{25/(9m-15)} .
\label{eq:tstar_scaling_main}
\end{equation}
Thus, \(t_\ast\) is a scaling estimate set by the fitted decay origin
and the initial distance of the upper tail from the Hinze scale. The critical value $m_c=5/3$ defines the marginal drift condition under which the Hinze scale and bubble population grow at the same rate. In classical freely decaying HIT,
including Saffman and Loitsyanskii--Batchelor decay, $m>m_c$, so the
stability boundary inevitably sweeps through the upper tail and suppresses
fragmentation.\\
For decay exponents relevant to Saffman and Loitsyanskii--Batchelor turbulence,
the Hinze scale grows faster than the characteristic bubble size and drives the population progressively into the sub-Hinze regime.  The mixed regime therefore does not merely delay coalescence: residual breakup initially sustains a large number density and enhances collision activity, while the growing Hinze scale simultaneously removes the breakable population. The resulting cascade-controlled transition is summarized in Table~\ref{tab:merged_scalings}, which lists the predicted scalings of $d$, $A$, $n$, $\Gamma$, $r$, $R$, $\mathcal{B}$, $d_H$, and $d/d_H$ in both regimes. These predictions are tested in Fig.~\ref{fig:d/d_H} and~\ref{fig:compensated_scaling} using the normalized observables $d_{32}/d_H$, $d_{99.8}/d_H$, $n$, $A$, and $r$, where $d_{32}$ is the Sauter mean diameter and $d_{99.8}$ characterizes the breakable upper tail of the bubble-size distribution. For the spatially developing duct-flow experiments, which are non  HIT, we use downstream distance as a convective decay coordinate, $t=x/U$, where $U$ is the bulk mean liquid velocity, and use the measured dissipation-decay exponent $m$ directly~\cite{kumar2026bubble,vivek_jfm}. This comparison tests whether the same
cascade-controlled scaling persists beyond ideal HIT, without assuming a
trajectory-level frozen-turbulence description for individual bubbles.

Fig~\ref{fig:d/d_H} examines these predictions by comparing the evolution of the normalized size ratios $d_{32}/d_H$ and $d_{99.8}/d_H$ across theory, DNS, and experiments. The DNS are performed for dilute bubbly HIT at $\phi$ of $0.5\%$ and $2\%$, ensuring that the flow remains in the one-way coupled regime while capturing variations in bubble population density. Experimental measurements are obtained in a strongly decaying turbulent duct flow over three $\phi$ ($= 0.5\%, 1\%, 2\%$) and three inlet turbulence intensities, spanning a wide range of initial Reynolds numbers and dissipation levels~\cite{kumar2026bubble,vivek_jfm}. Despite these variations in initial conditions, both our DNS and experiments exhibit a consistent collapse with our theoretical prediction when expressed in terms of $d_{32}/d_H$, $d_{99.8}/d_H$, and normalized time. In all cases with $m>5/3$, the ratio $d_{32}/d_H$ decreases monotonically, while the upper-tail ratio $d_{99.8}/d_H$ crosses below unity at the transition, confirming that the growth of the stability scale outpaces the growth of the dispersed phase and dynamically suppresses fragmentation.

 Notably, the early-time evolution exhibits a transient mixed regime in which breakup remains active, consistent with the initial presence of super-Hinze bubbles, before rapidly transitioning to a purely coalescence-dominated state. The agreement across idealized DNS HIT and strongly inhomogeneous duct-flow experiments demonstrates that the proposed scaling is not restricted to HIT, but captures the underlying physics of evolving turbulent cascade in realistic configurations. This comparison should be interpreted with care because bubble size, deformability, and wake interactions can modify turbulent transport and dissipation in non-HIT or bubble-driven configurations~\cite{Verschoof2016Drag,Ma2025Kolmogorov}. This unified collapse provides direct evidence that the evolution of dispersed-phase dynamics is governed by turbulence evolution rather than instantaneous stability criteria, establishing the central mechanism of the present study.
Fig.~\ref{fig:d/d_H} examines these predictions by comparing normalized forms of the size ratios $d_{32}/d_H$ and $d_{99.8}/d_H$ across theory, DNS, and experiments. The DNS are performed for dilute bubbly HIT at $\phi=0.5\%$ and $2\%$, where dispersed-phase feedback on the bulk turbulent decay remains weak while bubble population density varies. The DNS statistics are ensemble averages over two simulation realizations. The experimental campaign covers a strongly decaying turbulent duct flow over three $\phi$ ($=0.5\%,1\%,2\%$) and three inlet turbulence intensities, spanning a wide range of initial Reynolds numbers and dissipation levels~\cite{kumar2026bubble,vivek_jfm}; the figures show selected cases at $\phi=0.5\%$ and $2\%$. Despite these variations in initial conditions, the DNS and experiments exhibit decreasing normalized size ratios, consistent with the theoretical trends for $m>5/3$. For the DNS, the physical transition is identified primarily by depletion of the measurable super-Hinze tail and the condition $d_{99.8}<d_H$, with the absence of resolved breakup events providing supporting evidence [Fig.~\ref{fig:f_H}]. These observations indicate that growth of the stability scale outpaces growth of the dispersed phase and dynamically suppresses fragmentation.

Notably, the early-time evolution exhibits a transient mixed regime in which breakup remains active, consistent with the initial presence of super-Hinze bubbles, before rapidly transitioning to a purely coalescence-dominated state. The agreement in trends across idealized DNS HIT and strongly inhomogeneous duct-flow experiments supports the applicability of the proposed scaling beyond HIT. This comparison should be interpreted with care because bubble size, deformability, and wake interactions can modify turbulent transport and dissipation in non-HIT or bubble-driven configurations~\cite{Verschoof2016Drag,Ma2025Kolmogorov}. These observations support the interpretation that dispersed-phase evolution is governed by the evolving turbulent cascade and stability scale, rather than by an instantaneous stability criterion alone.

While Fig.~\ref{fig:d/d_H} illustrates the turbulence-driven suppression of fragmentation through the decrease of $d_{32}/d_H$, it does not directly address whether the underlying population dynamics exhibit universal behaviour across different flow conditions. The DNS and experiments are compared in a regime connected to recent measurements of finite-size bubble dispersion, pair dynamics, and deformation in isotropic turbulence~\cite{Mathai2018Dispersion,Tan2025PairDispersion,Xu2024Intermittency}. To examine this, Fig.~\ref{fig:compensated_scaling} considers the evolution of key dispersed-phase statistics: characteristic size, number density, interfacial area, and coalescence hazard, under a unified scaling framework. The time is normalized by the transition reference $t_\ast$, so that $\tau=t/t_\ast$, and the fitted series are compensated by their respective regime-dependent power laws. The DNS statistics are ensemble averages over two simulation realizations.

The theory predicts two dynamically distinct regimes. In the early-time mixed regime ($\tau<1$), where breakup and coalescence coexist, the theory predicts $d(t)\sim t^{\beta_t}$ with $\beta_t=(15+m)/25$, together with $n(t)\sim t^{-3\beta_t}$ and $A(t)\sim t^{-\beta_t}$, while the coalescence hazard scales as $r(t)\sim t^{-m/3-(2/3)\beta_t}$. As the turbulence decays, the transition near $\tau=1$ corresponds to depletion of the measurable super-Hinze tail, identified primarily by $d_{99.8}<d_H$ and supported by the absence of resolved breakup events. Beyond this point, breakup becomes negligible and the theory predicts $d(t)\sim t^{(3-m)/2}$, $n(t)\sim t^{-(9-3m)/2}$, $A(t)\sim t^{(m-3)/2}$, and a universal coalescence hazard $r(t)\sim t^{-1}$. The comparison is therefore framed in the spirit of decaying-turbulence studies in which large-scale structure and Reynolds-number effects constrain the observed decay and velocity statistics~\cite{Sinhuber2015Decay,Kuchler2023Universal}.

Using piecewise compensation based on regime-dependent empirical fits, Fig.~\ref{fig:compensated_scaling} shows that DNS and experimental data lie near unity across the void fractions and flow configurations shown. Separate compensation on the two sides of the transition follows the regime-dependent power-law description. Quantitative agreement with theory is assessed through the exponents. For the pure-coalescence DNS fits reported in Ref.~\cite{vivek_jfm}, the relative differences from theory, evaluated using the measured $m=2.48$ and $2.22$ at $\phi=0.5\%$ and $2\%$, are below $7\%$ for $d_{32}$, below $10\%$ for $A$, and $1\%$ and $6\%$ for $r$, respectively. These comparisons use the reported DNS fits, separately from the empirical fits used to compensate the figure. The experimental $d_{32}$ fits to the plotted input data differ from the prediction for $m=2.1$ by less than $8.7\%$, respectively.

\begin{center}
    \includegraphics[width=1\textwidth]{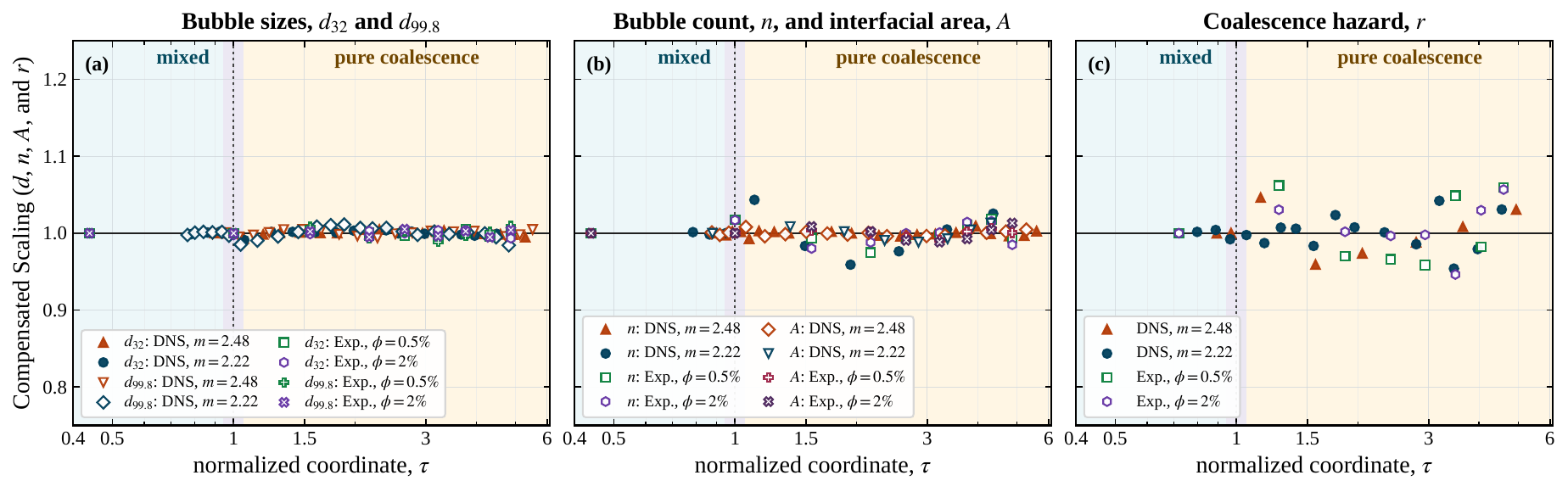}

    \nonfloatcaption{Dispersed-phase statistics displayed using piecewise compensated scaling. Time is normalized as $\tau=t/t_\ast$. Fitted series are compensated by separate empirical power laws in each regime, $Q/(C_i t^{\widehat{\alpha}_i})$; single-observation regimes use anchored theoretical exponents. For comparison, the theoretical scalings are: for $\tau<1$ (mixed coalescence-breakup regime), $\beta_t=(15+m)/25$ for the characteristic size, with $n\sim\tau^{-3\beta_t}$, $A\sim\tau^{-\beta_t}$, and $r\sim\tau^{-m/3-(2/3)\beta_t}$; for $\tau>1$ (pure coalescence regime), $d_{32}\sim\tau^\beta$ with $\beta=(3-m)/2$, $n\sim\tau^{-3\beta}$, $A\sim\tau^{-\beta}$, and $r\sim\tau^{-1}$.
    (a) Compensated characteristic and upper-tail diameters, $d_{32}$ and $d_{99.8}$. (b) Compensated number density $n$ and interfacial area $A$. (c) Compensated DNS coalescence hazard $r$, together with separately calculated experimental compensated hazard values.
    Symbols denote DNS with measured $m=2.48$ and $2.22$ at $\phi=0.5\%$ and $2\%$, respectively, and experiments with $m\simeq2.1$. The DNS plotting references $t_\ast=7$ and $8$ approximate the diagnosed transition times. Selected DNS markers are averaged over time intervals for clarity. The horizontal solid line denotes unity, and the vertical dotted line at $\tau=1$ separates the fitted regimes. The general theoretical scalings are summarized in Table~\ref{tab:merged_scalings}.}
    \label{fig:compensated_scaling}
\end{center}

In the theoretical description of the mixed regime, residual breakup maintains a large number density, so the characteristic bubble size grows with the enhanced transition exponent $\beta_t=(15+m)/25$, while $n(t)$, $A(t)$, and $r(t)$ follow the corresponding mixed-regime exponents. Once resolved breakup is suppressed, the same observables are predicted to follow the pure-coalescence exponents, with $d(t)\sim t^{(3-m)/2}$, $n(t)\sim t^{-3(3-m)/2}$, and $A(t)\sim t^{(m-3)/2}$. The DNS hazard exponents reported in Ref.~\cite{vivek_jfm}, $-0.99$ and $-1.06$, are close to the universal prediction $r(t)\sim t^{-1}$, consistent with late-time self-similarity once fragmentation is suppressed. Together with Fig.~\ref{fig:d/d_H}, these results support a dispersed-phase evolution governed by the time-dependent cascade that first sustains a collision-active mixed regime and then selects a pure-coalescence pathway. In this framework, the leading population exponents are predicted in terms of the measured decay exponent $m$, while the transition time depends on the initial upper-tail spread of the bubble-size distribution and on the fitted decay origin.

To conclude, a decaying turbulent cascade is not a passive background for dilute bubbly flow; through Hinze drift, it actively selects the fragmentation--aggregation pathway. For canonical decaying HIT, theory and DNS show universal evolution laws in both the transient mixed breakup--coalescence regime and the later pure-coalescence regime. Only the transition time retains memory of the initial distribution, set by the upper-tail distance \(\chi_i=d_{99.8}(t_i)/d_H(t_i)\), the decay exponent \(m\), and the fitted decay origin. For \(m>5/3\), including Saffman and Loitsyanskii--Batchelor decay, the Hinze scale outpaces bubble growth, sweeps through the super-Hinze tail, and dynamically suppresses fragmentation even when breakup is initially active. The population therefore crosses from a collision-active mixed regime to a self-sustaining sub-Hinze pure-coalescence state. Transient breakup is therefore not merely antagonistic to growth; it catalyzes coalescence by sustaining a larger number density and stronger collision activity before the breakable tail disappears. After this shutdown, the population follows pure-coalescence scaling with a universal per-bubble coalescence hazard \(r(t)\sim t^{-1}\). The collapse of high-Reynolds-number duct-flow experiments when normalized by the measured decay rate shows that this Hinze-drift mechanism persists beyond ideal HIT. Together, these results replace an instantaneous Weber-number view of breakup with a non-equilibrium stability framework governed by the relative evolution of the turbulent cascade and the bubble-size distribution, supplying predictive scaling constraints for population-balance, interfacial-area, and closure models of turbulent fragmentation--aggregation\cite{ChanJohnsonMoin2021CascadePart1,ChanJohnsonMoinUrzay2021CascadePart2,QiTanCorbittUrbanikSalibindlaNi2022SmallEddies,RivierePerrard2025BreakupProbability,kumar2026bubble}.

Acknowledgments -- Authors thank Alina Baza for the help with plotting and figures. This work was supported in part by the Advanced Research Projects Agency--Energy, U.S. Department of Energy (DOE) under Award Number DE-AR0001587, and Office of Critical Minerals and Energy Innovation (Former EERE), US DOE under Award Number DE-EE0009396. \\S.~S.~J acknowledges support from the donors of the ACS Petroleum Research Fund through Doctoral New Investigator Grant 69196-DNI9 and the NSF CAREER award with grant number GR00037347. 
The authors also acknowledge computing resources provided through the U.S. Department of Energy 2024 and 2025 ALCC awards (TUR147 and BubbleLaden). The Argonne Leadership Computing Facility at Argonne National Laboratory is supported by the Office of Science of the U.S. Department of Energy under Contract No. DE-AC02-06CH11357. The Oak Ridge Leadership Computing Facility at Oak Ridge National Laboratory is supported by the Office of Science of the U.S. Department of Energy under Contract No. DE-AC05-00OR22725.
 \\
Supplementary materials -- the detailed derivation is provided in
\textbf{\texttt{supplementary.pdf}}, together with the supplementary movie
\textbf{\texttt{movie.avi}}.

\bibliography{cleaned_references}
\end{document}